\documentstyle[aps,twocolumn]{revtex}
\input{psfig}

\def\be{\begin{equation}}
\def\ee{\end{equation}}
\def\beq{\begin{eqnarray}}
\def\eeq{\end{eqnarray}}

\begin{document}
\draft
\wideabs{
\title{Edge State Transport of Separately Contacted Bilayer Systems\\ in the
Fractional Quantum Hall Regime}

\author
{Daijiro {\sc Yoshioka} and Kentaro {\sc Nomura}}

\address{
Department of Basic Science, The University of Tokyo\\
3-8-1 Komaba, Meguro, Tokyo 153-8902, Japan
}
\maketitle
\begin{abstract}
Hall and diagonal resistances of bilayer fractional quantum Hall systems are
discussed theoretically.
The bilayers have electrodes attached separately to each layer.
They are assumed to be coupled weakly by interlayer tunneling, while the
interlayer Coulomb interaction is negligibly small.
It is shown that source-drain voltage dependence of the resistances reflects
the Luttinger liquid parameter of the edge state.
\end{abstract}
\pacs{73.40.-c, 73.40.Gk,73.40.Hm, FQHE, bilayer, tunneling, Luttinger liquid
}
}

Recently it has become possible to fabricate bilayer systems where electrodes
are attached separately to each layers\cite{eisen}.
Most experimental and theoretical investigations have been done for the
phenomenon called as Coulomb drag in these systems, which occurs only when the
system is in the compressible phase.
However, even in the incompressible states non-trivial transport phenomena are
expected to occur if the two-layers are coupled through tunneling.
Actually, in the integer quantum Hall regime deviation from the quantized Hall
value and finite dissipation have been observed experimentally\cite{ohno}, and
analyzed theoretically\cite{ym,ym1}.
The purpose of the present paper is to investigate the consequences of the
tunneling for weakly coupled bilayer systems in the fractional quantum Hall
regime.

For the bilayer system, where current is supplied to only one of the layer, the
two layers are out of equilibrium, especially at the edges.
The tunneling cause the two layers to approach to the equilibrium.
It is known that the tunneling in the bulk is suppressed as long as the voltage
difference between the two-layers is small\cite{eisen1}.
Therefore we assume that the tunneling occurs only between the chiral Luttinger
liquids realized at the sample edge.

However, even at the edges, it has been shown that for the principal quantum
Hall states realized at filling factor $1/q$ $(q > 1)$ the tunneling is
irrelevant and suppressed in contrast to the integer quantum Hall (IQH) case,
where it is relevant\cite{ny}.
For the case of FQH, The irrelevancy of the tunneling gives power law
dependence of the tunneling current on the voltage difference or on the
temperature\cite{grayson}.
The power is given by the Luttinger liquid parameter, which is nothing but the
filling factor of the system.
This theory suggests new experiment to investigate the Luttinger liquid
property of the fractional quantum Hall edges.

Following \cite{ym} we introduce local effective chemical potential
$\mu_\sigma(x)$ to describe the current flowing in the $\sigma$-layer at
coordinate $x$ along the edge, where $\sigma=\pm$ is the layer index.
The introduction of the effective chemical potential does not necessarily mean
that each layer is in thermal equilibrium\cite{note1}.
The current in each layer is given by
\be
I_\sigma(x) = {{\nu e}\over{h}} [\mu_\sigma(x) - \epsilon_0],
\ee
where $\epsilon_0$ is a common origin to measure energy and defines the edge
current.
Due to the chirality of the edges the sum of the current $I_+(x)+I_-(x)$ is
conserved along the edges both from source to drain and from drain to source.
However, the tunneling current causes the change in the current in each layers.
Since the electrons flow at a constant velocity along the edges, the temporal
evolution of the difference in the chemical potential is projected to the
spatial evolution of the chemical potential.
The spatial evolution should be described by Boltzman type differential
equation as follows
\be
{{{\rm d}\mu_\sigma(x)}\over{{\rm d}x}} = - {{1}\over{\xi}}
[\mu_\sigma(x) - \mu_{-\sigma}(x)]^\lambda.
\ee
In this equation $\lambda$ is given by the Luttinger parameter of the edge
state.
For the standard theory of the edge state in the principal FQH state at
$\nu=1/q$, $\lambda = 2q-1$ \cite{wen}.
However, we can consider $\lambda$ as a parameter to be determined
experimentally.
The parameter $\xi$, which has the meaning of the relaxation length for the
IQH, is inversely proportional to the tunneling probability.
We solve this equation under the condition that the total current through the
sample is $I$, and only the minus layer is connected to the source and drain
electrodes.
We assume that the current is injected ideally without reflection into the
upper edge of the minus layer from the source at $x=0$, and drained at $x=L$,
namely
$L$ is the sample length.
Thus the boundary condition for the minus layer is $\mu_-(0+)=\mu_{\rm S}$, and
$\mu_-(L-)=\mu_{\rm D}+ (h/\nu e)I$.
Similarly boundary condition for the lower edge of the minus layer is
$\mu_-(L+)=\mu_{\rm D}$, and $\mu_-(2L-)=\mu_{\rm S} - (h/\nu e)I$, where
neglecting the width of the electrodes, the edge state extends from the drain
at $x=L$ to the source at $x=2L \equiv 0$.
On the other hand, since the upper layer is not connected to the current
electrodes, the condition for $\mu_+(x)$ is $\mu_+(0-)=\mu_+(0+)$ and
$\mu_+(L-)=\mu_+(L+)$.

To solve the equations we introduce an auxiliary variable $\zeta$ such that
\beq
(\mu_{\rm S} - \mu_{\rm D})\zeta &=& \mu_-(0+) - \mu_-(L-)\cr
 &=& -[\mu_+(0) - \mu_+(L)] \cr
&=& \mu_-(L+) - \mu_-(0-).
\eeq
This $\zeta$ is determined by the following transcendental equation:
\beq
(1-3\zeta)^{1-\lambda} &-& (1+\zeta)^{1-\lambda}\cr
 &=& 2(\lambda -1) {{L}\over{\xi}}
[{{1}\over{2}}(\mu_{\rm S} - \mu_{\rm D})]^{\lambda-1}.
\eeq
Once $\zeta$ is determined, $\mu_\pm (x)$ is given:
Namely for $0 < x < L$,
\beq
\mu_\pm(x) &=& {{1}\over{4}}[(3-\zeta)\mu_{\rm S} + (1+\zeta)\mu_{\rm D}]\cr
&\mp& {{1}\over{2}}[\{ {{1+\zeta}\over{2}}
(\mu_{\rm S}-\mu_{\rm D})\}^{1-\lambda}\cr
& & + {{2(\lambda -1)}\over{\xi}} x]^{1/(1-\lambda)},
\label{eq:1}
\eeq
and for $L < x <2L$,
\beq
\mu_\pm(x) &=& {{1}\over{4}}[(1+\zeta)\mu_{\rm S} + (3-\zeta)\mu_{\rm D}]\cr
&\pm& {{1}\over{2}}[\{ {{1+\zeta}\over{2}}
(\mu_{\rm S}-\mu_{\rm D})\}^{1-\lambda}\cr
  & & + {{2(\lambda -1)}\over{\xi}}(x-L)]^{1/(1-\lambda)}.
\label{eq:2}
\eeq
It is easily verified that $\mu_\pm(x)$ satisfies the differential equation
with the proper boundary conditions.

Now, the current through the sample is given as follows:
\be
I=\nu{{e}\over{h}} (1-\zeta)(\mu_{\rm S} - \mu_{\rm D}).
\ee
Therefore the source-drain conductance is given by
\be
G={{\nu e^2}\over{h}} (1-\zeta).
\ee
Hall and diagonal resistances when the probe electrodes are attached to the
minus layer at $x=a$, $x=L-a$, and $x=2L-a$ are calculated from
eqs.(\ref{eq:1}) and (\ref{eq:2}):
\beq
R_{\rm H} &\equiv& {{ \mu_-(a) - \mu_-(2L-a)}\over{eI}} \cr
&=& ({{h}\over{\nu e^2}})
\{ {{1}\over{2}} +
{{1}\over{4(1-\zeta)}}[{{L-a}\over{L}}
(1+\zeta)^{1-\lambda} \cr
& &+ {{a}\over{L}}(1-3\zeta)^{1-\lambda}]^{1/(1-\lambda)}\cr
&+&
{{1}\over{4(1-\zeta)}}[{{a}\over{L}}
(1+\zeta)^{1-\lambda} \cr
& & + {{L-a}\over{L}}(1-3\zeta)^{1-\lambda}]^{1/(1-\lambda)}
\},
\eeq
and
\vskip 1cm
\beq
R_{xx} &\equiv& {{\mu_-(a) - \mu_-(L-a)}\over{eI}}\cr
&=&({{h}\over{\nu e^2}}) {{1}\over{4(1-\zeta)}}
\{ [{{L-a}\over{L}} (1+\zeta)^{1-\lambda}\cr
& &+ {{a}\over{L}}(1-3\zeta)^{1-\lambda}]^{1/(1-\lambda)}\cr
&-&
[{{a}\over{L}}(1+\zeta)^{1-\lambda}\cr
& &+ {{L-a}\over{L}}(1-3\zeta)^{1-\lambda}]^{1/(1-\lambda)}
\}.
\eeq

These results are shown in Figs. 1 and 2.
First in Fig.1 the normalized conductance $G/(\nu e^2/h) = (1-\zeta)$ is
plotted as a function of $(L/\xi) [(\mu_{\rm S} - \mu_{\rm D})/2]^{\lambda-1}$.
As the tunneling current increases either as the length $L$ or $(\mu_{\rm S} -
\mu_{\rm D})$ increase or $\xi$ decreases, the source-drain conductance
decreases.
This is due to the increase of the back-scattered current through the plus
layer.
A peculiarity of the FQH case is that it depends on $(\mu_{\rm S} - \mu_{\rm
D})$.
The conductance tends to $(2/3)(\nu e^2/h)$ as the tunneling current increases.

In Fig.2 the Hall and longitudinal resistances, which are normalized by $h/\nu
e^2$, are plotted for the choice of $a=0.2L$.
Effect on the longitudinal resistance is smaller in the fractional case.
The Hall resistance tends to 1/2 of the isolated layer case in the large
tunneling current limit, while $R_{xx}$ tends to zero similarly to the IQH
case.

The results in this paper show that if $(\mu_{\rm S} - \mu_{\rm D})$-dependence
of the various quantity is measured for tunnel-coupled bilayer systems, we can
deduce the parameter $\lambda$, or the Luttinger parameter of the edge state.
Of course such an experiment should be done at low temperature and for small
source-drain voltage, since tunneling in the bulk must be negligible.
We hope such an experiment will be done in the near future.
This work is supported by Grant-in-Aid for Scientific Research (C) 10640301.

%
%
\begin{figure}
\psfig{figure=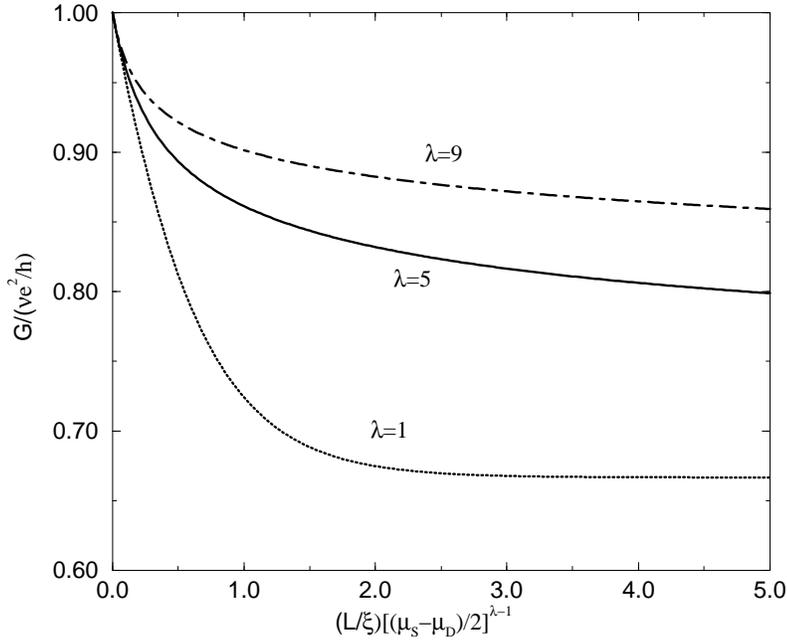,height=9cm}
\caption{Two terminal conductance $G$ scaled by $\nu e^2/h$ is shown
as a function of tunneling strength $(L/\xi)[(\mu_{\rm S} -\mu_{\rm D})
/2]^{\lambda -1}$ for $\lambda = 1$, 5, and 9 by dotted, solid and dot-dashed
lines, respectively.}
\label{fig:1}
\end{figure}

%
%
\begin{figure}
\psfig{figure=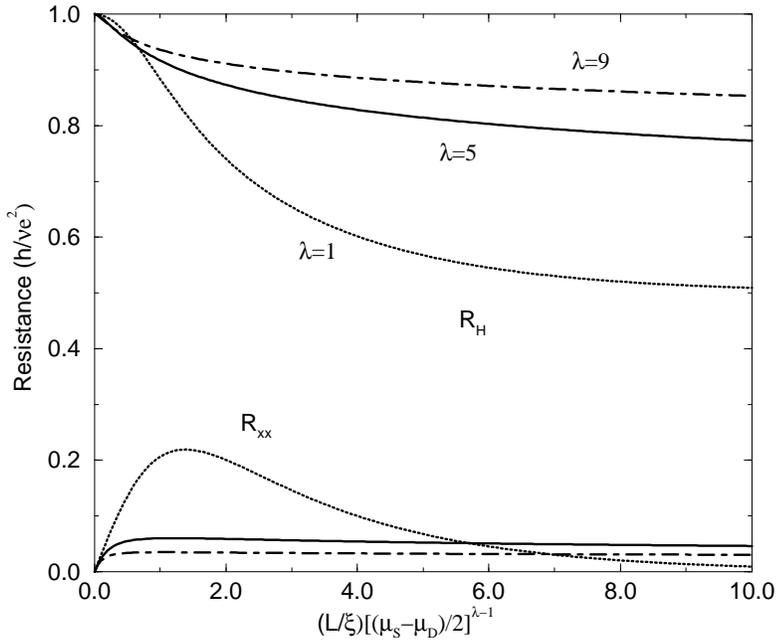,height=9cm}
\caption{The diagonal and Hall resistances $R_{xx}$ and $R_{\rm H}$
scaled by $h/\nu e^2$ are shown
as a function of tunneling strength $(L/\xi)[(\mu_{\rm S} -\mu_{\rm D})
/2]^{\lambda -1}$ for $\lambda = 1$, 5, and 9 by dotted, solid and dot-dashed
lines, respectively.
The length of the sample is $L$ and voltage probes are attached at 0.2$L$
from the source and drain.}
\label{fig:2}
\end{figure}


\begin{references}
\bibitem{note1} If the interlayer relaxation is faster than the intralayer
relaxation, each edge will be in thermal equilibrium, and the effective
chemical potential is nothing but the true chemical potential.
\end{references}
\end{document}